\documentclass[
    ,final            
  ]
  {aipproc}

\layoutstyle{6x9}

\begin{document}

\title{Physics Updates from HERMES}


\author{Morgan J. Murray on behalf of the H{\sc ermes } Collaboration}{
  address={Rm. 514b, Dept. of Physics \& Astronomy, University of
    Glasgow, Glasgow, G12 8QQ, Scotland}
}

\begin{abstract}
The H{\sc ermes} collaboration presents two sets of recent results:
the first related to an extraction of the $g_2$ nucleon structure
function determined for DIS on a polarised target and the second
related to the measurement of asymmetries related to the deeply
virtual Compton scattering process that can be used to discover
information on generalised parton distributions and hence nucleon structure.
\end{abstract}

\maketitle


\section{The HERMES Experiment}
The H{\sc ermes} experiment was a fixed-target experiment on the H{\sc
  era} ring in Hamburg, Germany. The experiment ran from 1995-2007 and
used the electron/positron beam made available by the H{\sc era}
accelerator. A variety of polarised targets were used; the targets
under consideration in these proceedings were transversely polarised
(w.r.t. the beam direction) and polarisation averaged proton targets.

\section{Extraction of the $g_2$ structure function}

\subsection{Experimental Details}

The $F_1$, $F_2$ and $g_1$ structure functions are well-known
quantities in the field of nucleon structure physics. Much less is
known about the $g_2$ structure function, which (unlike the
aforementioned functions) has no intuitively-graspable interpretation
as a probability density. Compared to its brother-functions, $g_2$
(and its related asymmetry $A_2$) has
undergone relatively little experimental scrutiny. There have been
some prior experimental extractions of $g_2$ at SLAC and CERN~\cite{slacg21,slacg22,smcg2}. The
former facility has observed evidence of unexpected potential twist-3 behaviour,
i.e. they deviate from the twist-3 behaviour calculated using the
Wandzura-Wilczek approximation.

The $A_2$ asymmetry and $g_2$ distribution were extracted at H{\sc
  ermes} from data taken in 2003-2005~\cite{hermesg2}, when a transversely polarised
Hydrogen gas target was used in the experiment. The target has an
average polarisation of 78\% and the electron/positron beam was
polarised to approximately 34\%. The applied kinematic boundaries for the
accumulated data set were
$0.18\,\textrm{GeV}^2<Q^2<20\,\textrm{GeV}^2$, $W>1.8\,\textrm{GeV}$,
$0.004<x<0.9$ and $0.10<y<0.91$. The data were corrected for the
charge-symmetric $e^+e^-$ background, which
amounted in total to about 1.8\% of the events, reaching
the largest contribution of about 14\% at small values of
x. The asymmetry 
\begin{equation}
A_{LT}(x,Q^2,\phi,h_\ell) = h_\ell\frac{N^{h_\ell
    \Uparrow}(x,Q^2,\phi)\mathcal{L}^{h_\ell\Downarrow} - N^{h_\ell
    \Downarrow}(x,Q^2,\phi)\mathcal{L}^{h_\ell\Uparrow}}{N^{h_\ell
    \Uparrow}(x,Q^2,\phi)\mathcal{L}_p^{h_\ell\Downarrow} + N^{h_\ell
    \Downarrow}(x,Q^2,\phi)\mathcal{L}_p^{h_\ell\Uparrow}}
\end{equation}
was measured in bins of $x$, $Q^2$, and the azimuthal scattering angle
$\phi$ where $N^{h_\ell
    \Uparrow(\Downarrow)}(x,Q^2,\phi)$ is the number of scattered leptons in each
bin for the case of an incident lepton with helicity $h_\ell$ when the direction
of the proton spin points up (down). The quantities
$\mathcal{L}^{h_\ell\Uparrow(\Downarrow)}$ and $\mathcal{L}_p^{h_\ell\Uparrow(\Downarrow)}$
are the corresponding integrated luminosities and the in-
tegrated luminosities weighted with the absolute value of
the beam and target polarization product, respectively. The measured
asymmetties are corrected for radiative and instrumental smearing
effects. The resultant data in each ($x$,$Q^2$)-bin were fit with a
functional form of $A_T\cos\phi$ and $A_2$ and $g_2$ were calculated
from $A_T$ with knowledge of the $F_2$ and $g_1$ structure functions
and $R(x,Q^2)$, the ratio of longitudinal to transverse virtual-photon absorption cross sections.

\subsubsection{Result}
\begin{figure}
  \includegraphics[width=\textwidth]{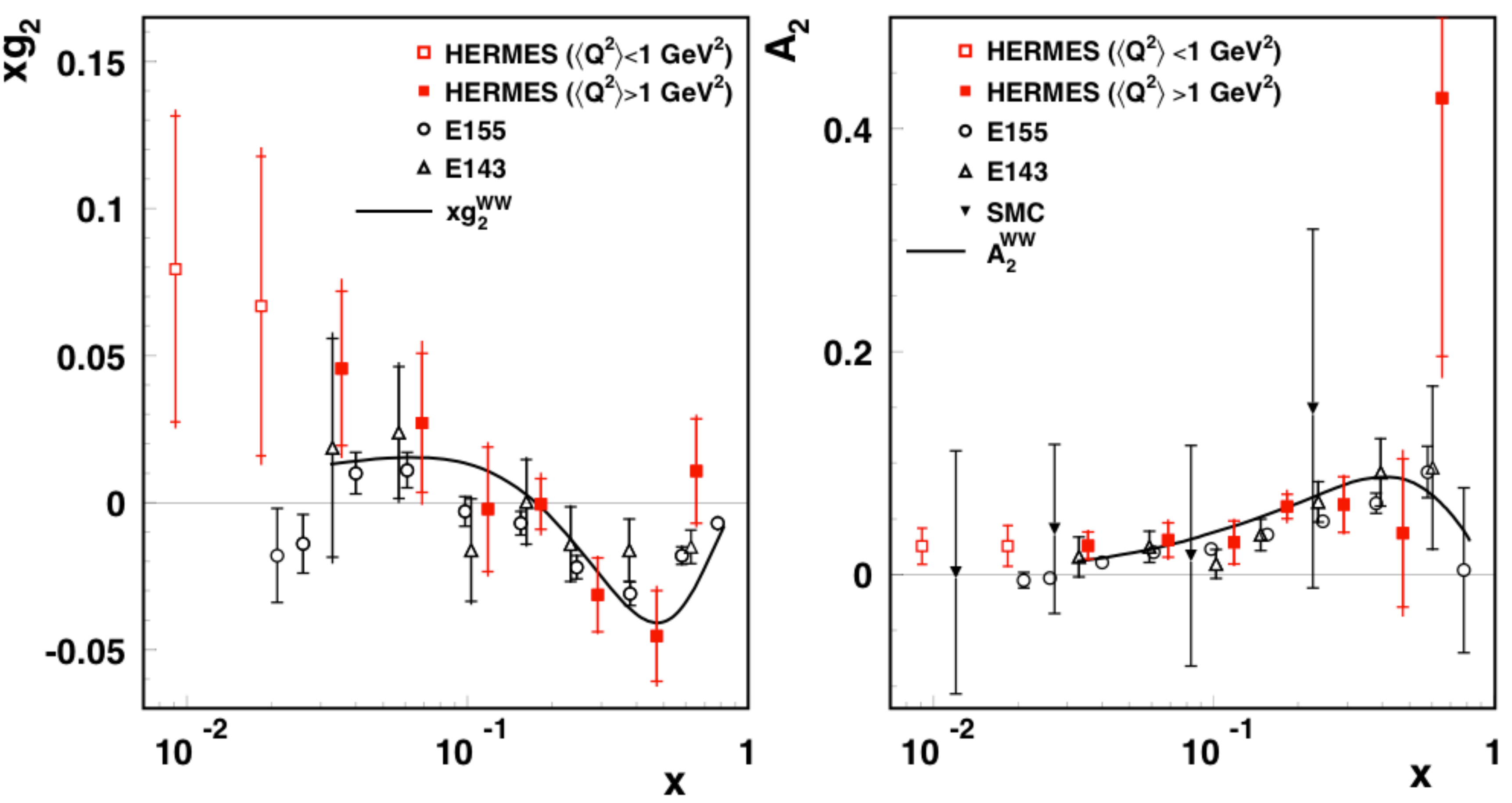}
  \caption{Values of $g_2$ (left) and $A_2$ (right) extracted at H{\sc
      ermes} (red squares) in two bins of $Q^2$ and projected along
    $x$. The results are
    compared to previous measurements taken at SLAC and CERN and a
    Wandzura-Wilczek calculation of twist-3 effects. Inner error bars
    are the statistical uncertainties; outer bars are the statistical
    and systematic uncertainties added in quadrature.}
  \label{fig:a2g2}
\end{figure}
The resultant values are shown in figure \ref{fig:a2g2}. The results
from H{\sc ermes} are split into two different bins in $Q^2$ and
projected along $x$. The H{\sc ermes} data is consistent with previous
measurements, but not sufficiently precise to make conclusions about
potential deviations from twist-2 behaviour. The results for $g_2$ are
consistent with the Cottingham-Burkardt sum rule~\cite{cbrule}, which states that
for sufficiently large $Q^2$ the integral of $g_2$ over $x$ should be zero.

\section{Exclusive Physics}

\begin{figure}
  \includegraphics[width=1.1\textwidth]{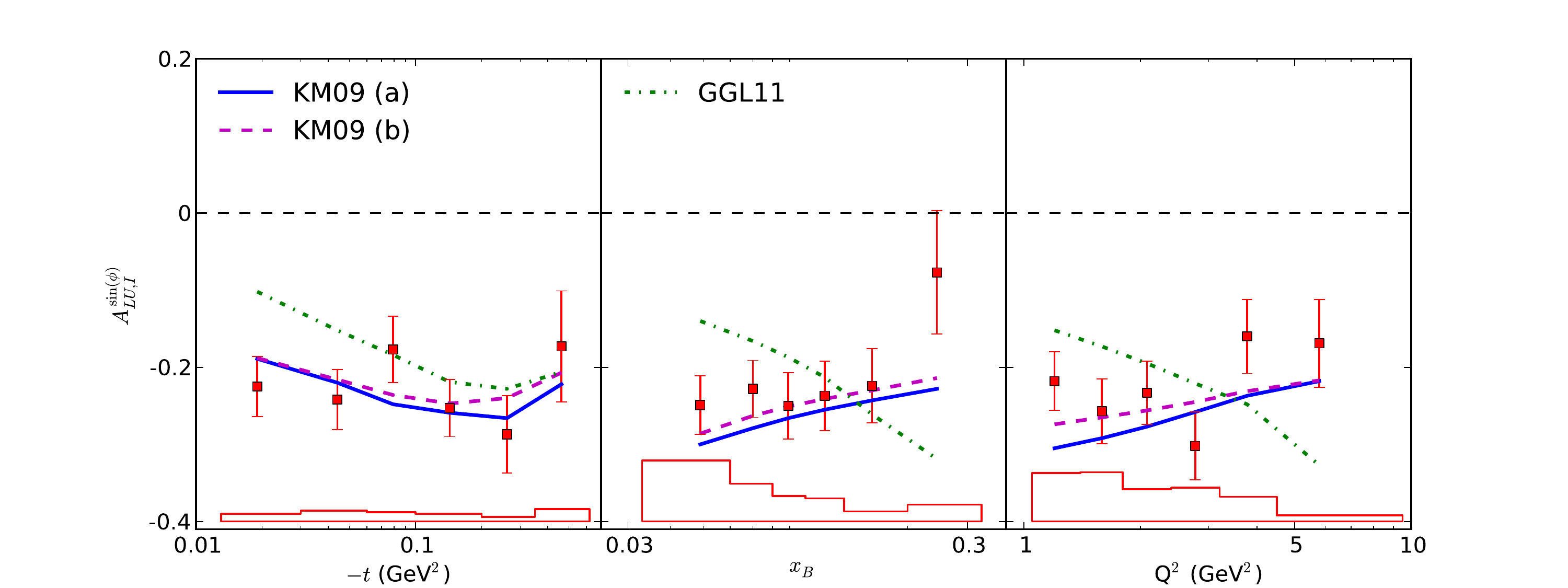}
  \caption{The interference contribution to the first
    harmonic of the beam helicity asymmetry projected in six bins
    along the three kinematic dimensions $-t$, $x_{\mathrm{B}}$ and
    $Q^2$. This measurement can be used to constrain  a part of the
    GPD $H$, the generalised parton distribution that reduces to the
    parton distribution $q(x)$ in its forward limit. Error bars are
    the statistical uncertainties; error bands are the systematic
    uncertainties. The curves on the figure come from \cite{km} and \cite{ggl}.}
  \label{fig:bsa}
\end{figure}

\begin{figure}
  \includegraphics[width=1.1\textwidth]{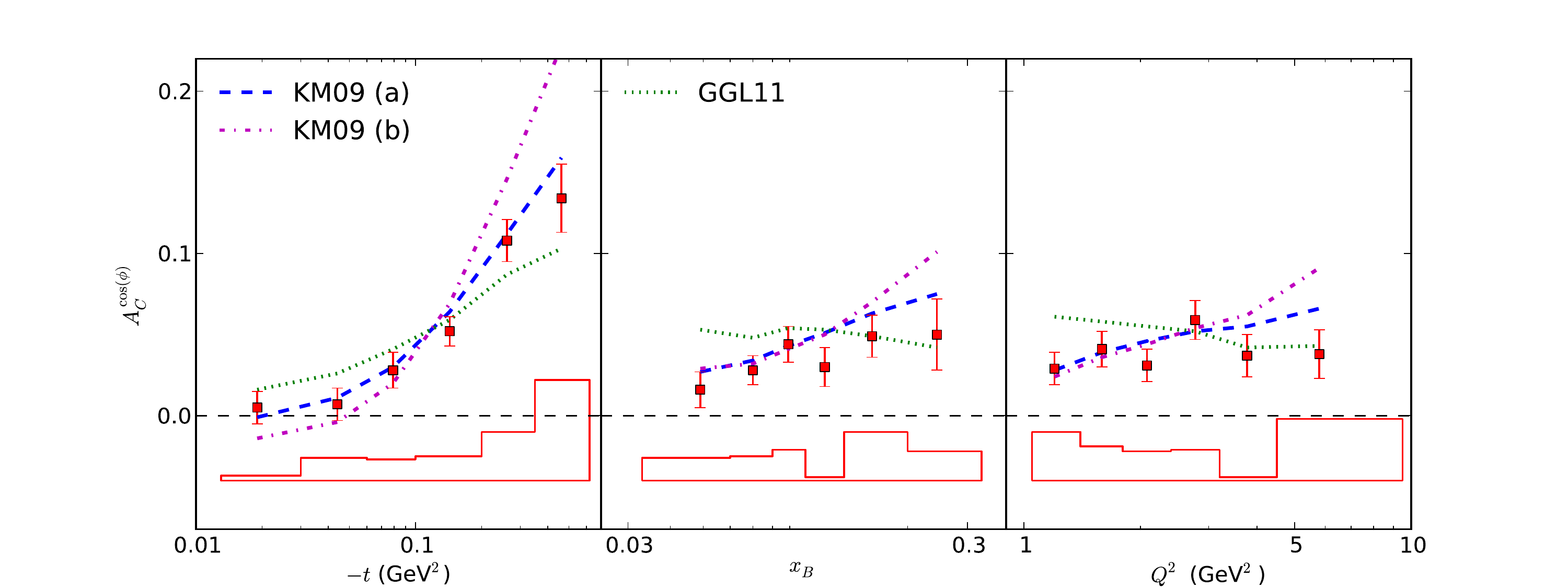}
  \caption{The DVCS and interference contributions to the first
    harmonic of the beam charge asymmetry. This measurement can be
    used to constrain a part of the GPD $H$, the generalised parton distribution
    that reduces to the parton distribution $q(x)$ in its forward
    limit. The curves on the figure come from \cite{km} and \cite{ggl}.}
  \label{fig:bca}
\end{figure}

\begin{figure}
  \includegraphics[width=\textwidth]{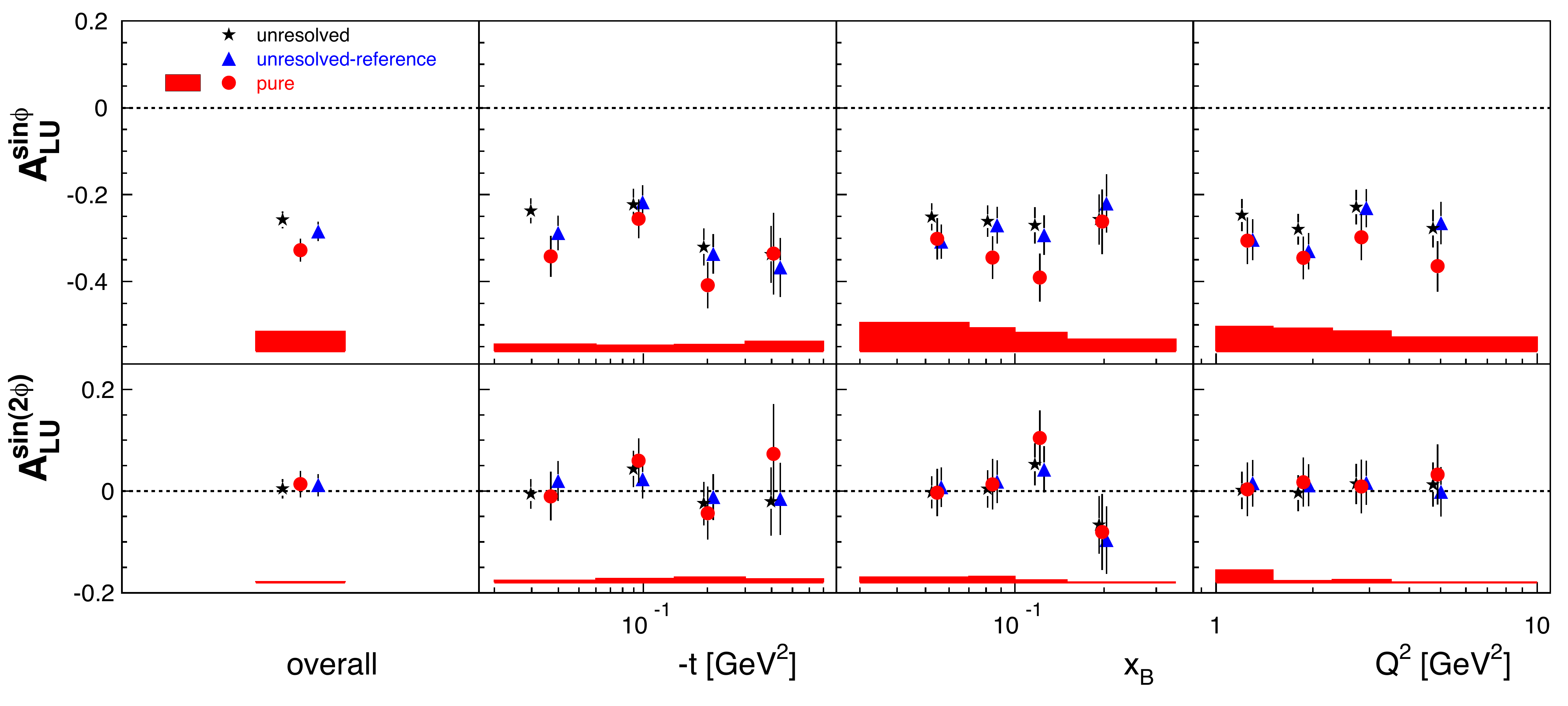}
  \caption{The beam helicity asymmetry measured using a missing-mass
    event selection technique in the original H{\sc
      ermes} acceptance (black), in the H{\sc ermes} acceptance as
    modified by the installation of the recoil detector (blue) and
    with a highly pure exclusive event selection technique (red). The
    first harmonic shows that a pure sample has a slightly larger
    asymmetry value at the larger $-t$ values associated with high
    contamination in the missing-mass selected sample by resonsant
    events. The second harmonic is compatible with zero. }
  \label{fig:recoil}
\end{figure}

\begin{figure}
  \includegraphics[width=\textwidth]{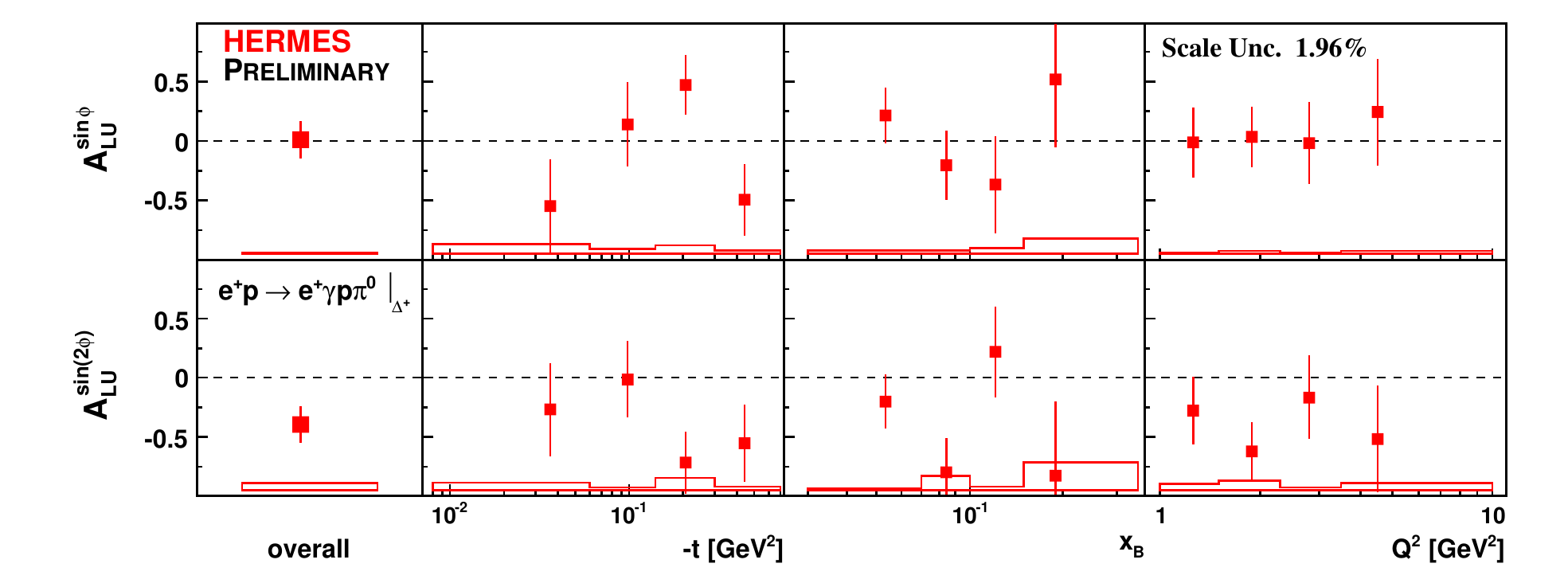}
  \caption{The beam helicity asymmetry associated with the process
    $ep\rightarrow e\Delta\gamma\rightarrow ep\pi^0\gamma$, which is
    partly responsible for a 12\% contamination of the missing-mass
    selected data sample. The amplitude of the asymmetry is compatible
  with zero.}
  \label{fig:assoc}
\end{figure}

Knowledge of nucleon structure can be expanded by considering generalised
parton distributions (GPDs); access to GPDs can be achieved by measuring
particle production in scattering processes where the target nucleon
remains intact. The simplest process to measure is deeply virtual
Compton Scattering (DVCS), where a photon is produced by a parton from
the nucleon~\cite{hermesBSABCA1,hermesBSABCA2,hermesALT,hermesLTSA,hermesTTSA}. It is also possible to interpret measurements of produced
mesons in the GPD framework, e.g.~\cite{hermesMESON1,hermesMESON2,hermesMESON3,hermesMESON4}. The H{\sc ermes} experiment
recently produced DVCS measurements from the entire 1995-2007 data
set, where beam helicity and charge azimuthal asymmetries in the produced
particle distributions are measured, see figures~\ref{fig:bsa}
and~\ref{fig:bca} respectively. The H{\sc ermes} experiment can
distinguish the pure DVCS contribution to the asymmetries from the
contribution due to the interference with the competing Bethe-Heitler
process due to the presence of both beam charges in the data
set. These measurements are taken with a missing-mass selection
technique, where the target nucleon scatters outside the H{\sc ermes} geometric acceptance and must be
reconstructed from measurements of the produced photon and scattered
lepton. This event selection technique allows a certain amount of
background into the data sample, mostly from events involving a
resonant state of the target nucleon. The collaboration has also
published measurements from an kinematically completely recostructed data sample~\cite{hermesRECOIL} that was taken with additional
experimental equipment---the beam helicity asymmetry measured using an
exclusive event selection technique is shown in
figure~\ref{fig:recoil}. Unfortunately, there is only a single beam charge
available for this additional dataset, so it is not possible to
distinguish between the contributions from the interference and
$\vert\textrm{DVCS}\vert^2$ terms to the
asymmetry. It is, however, possible to make a first extraction of the
beam helicity asymmetry associated with the resonant events that
contaminate the missing-mass selected data sample. The beam helicity
asymmetry for the process $ep\rightarrow e\Delta\gamma\rightarrow
ep\pi^0\gamma$ is shown in figure~\ref{fig:assoc}.

\begin{theacknowledgments}
 We gratefully acknowledge the D{\sc esy} management for its support, the staff
at D{\sc esy} and the collaborating institutions for their significant effort,
and our national funding agencies and the EU FP7 (HadronPhysics2, Grant
Agreement number 227431) for financial support.
\end{theacknowledgments}



\bibliographystyle{aipproc}   




\end{document}